\documentclass[11pt]{article}
\hoffset = - 2 truecm 
\def\beq{\begin{equation}}   \def\eeq{\end{equation}}
\def\bea{\begin{eqnarray}}  \def\eea{\end{eqnarray}} \def\nn{\nonumber}
\def\noi{\noindent} \def\beeq{\begin{eqnarray}}
\def\eeeq{\end{eqnarray}}
\def\lsim{\raise0.3ex\hbox{$<$\kern-0.75em\raise-1.1ex\hbox{$\sim$}}}
\def\gsim{\raise0.3ex\hbox{$>$\kern-0.75em\raise-1.1ex\hbox{$\sim$}}}

\usepackage{graphicx}

\usepackage{epsfig}

\begin{document}
\pagestyle{plain}
\baselineskip 18pt
\begin{titlepage}
 \begin{flushright}

           \today\\ 
	   LPT-Orsay-14-85 \\ 
	LAPTH-229/14 \\
\end{flushright}

\vspace{1.cm}

\begin{center}

\vbox to 1 truecm {}



{\large \bf Photon-Jet cross sections in Deep-Inelastic Scattering}
\vskip 2 truecm

{\bf P. Aurenche$^{1,a}$ and M. Fontannaz$^{2,b}$} \vskip 3 truemm

{\it $^1$ LAPTh, Universit\'e de Savoie, CNRS
\\ BP 110, Chemin de Bellevue, 74941 Annecy-le-Vieux Cedex, France}

\vskip 3 truemm
{\it $^2$ Laboratoire de Physique Th\'eorique, UMR 8627 du CNRS,\\
Universit\'e Paris-Sud, B\^atiment 210, 91405 Orsay Cedex, France}
\vskip 2 truecm

\begin{abstract}
We present the complete next-to-leading order calculation of isolated prompt photon production in association with a jet in deep-inelastic scattering. The calculation involves, direct, resolved and fragmentation contributions. It is shown that defining the transverse momenta in the proton virtual-photon frame (CM$^*$), as usually done, or in the laboratory frame (LAB), as done in some experiments, is not equivalent and leads to important differences concerning the perturbative approach. In fact, using the latter frame may preclude, under certain conditions, the calculation of the next-to-leading order correction to the important resolved component. A comparaison with the latest ZEUS data is performed and good agreement is found in the perturbatively stable regions.
\end{abstract} 
\end{center}

\vfill
$^a$ e-mail: patrick.aurenche@lapth.cnrs.fr\par
$^b$ e-mail: michel.fontannaz@th.u-psud.fr\par
\end{titlepage}
\newpage
\section{Introduction}
\hspace*{\parindent} 
Large transverse momentum phenomena in deep inelastic scattering reactions have been extensively studied by the H1 and ZEUS collaborations at HERA. Large-$p_\bot$ particle and jet spectra have been measured and compared to next-to-leading order (NLO) QCD calculations. Large-$p_\bot$ photons have also been observed, in an inclusive way \cite{a,b} or in correlation with a hadronic jet \cite{a,c}. This latter reaction has been the subject of theoretical studies some ten years ago \cite{d,e}. The recent data from ZEUS~\cite{c} lead us to extend these studies and to compare the complete NLO QCD results with the $\gamma$-jet cross sections.

In principle, prompt photon production in deep-inelastic scattering (DIS) is a very simple process~: it goes via the Compton scattering of a virtual photon on a quark: $\gamma^* + q \to \gamma + q$ and requires only the knowledge of the distribution function of a quark in the proton. Including higher-order (HO) corrections considerably complicates the picture and new objects have to be introduced.
For example, in the scattering $\gamma^*+ g \rightarrow q + \bar q + \gamma$, the $q\bar q$ pair may be produced quasi-collinearly to the virtual photon, one of the parton in the pair being then scattered at large $p_\perp$: this configuration generates the virtual photon structure function (resolved photon) at lowest order associated to a large logarithm. It is then necessary to resum such large logarithms and introduce the all order photon structure function.
Furthermore, in the above process or in $\gamma^*+ q \rightarrow q +  g + \gamma$, the final photon may be produced collinearly to a final state quark or antiquark (bremsstrahlung) leading to a large logarithmic enhancement, thus generating the photon fragmentation fonction. Thus one is lead to distinguish four types of processes, all contributing already at leading order (LO): the direct-direct (d-d) one where both the virtual photon and the final real photon couple directly to the hard sub-process; the resolved-direct (r-d) where the virtual photon couples to the hard subprocess through its hadronic (perturbative or non perturbative) components; the direct-fragmented (d-f) and the resolved-fragmented (r-f) ones where the final photon appears as a fragment of a jet unlike in the previous two cases where it is isolated.  At HERA, all four processes corresponding to four topologies have essentially the same order of magnitude \cite{g}. However when dealing with isolated photon, the isolation criteria necessary to suppress the background from $\pi^0 \to \gamma\gamma$, considerably reduces the fragmentation components d-f and r-f.

The above discussion on the four topologies is valid as long as we can define a virtual photon structure function resumming all the large logarithms $\ln \left ( {p_\bot^2 + Q^2 \over Q^2} \right )$ \cite{f} where $p_\bot$ is a characteristic transverse momentum of the reaction (for instance that of the observed photon in the proton virtual-photon center-of-mass frame) and $Q^2$ the initial photon virtuality. These terms appear in the calculation of HO  corrections to the Born direct cross sections. If $p_\bot$ is not large enough ($p_\bot^2 \ \lsim\ Q^2$) it is of course not useful to subtract these logarithms from the direct HO corrections in order to resum them in the virtual photon structure function. On the other hand for $p_\bot^2 \gg Q^2$ this approach is useful~: indeed in this case the resolved cross sections have the same structure as a hadronic cross section involving two parton distributions convoluted with hard subprocesses. HO corrections are known, they are large and can be easily implemented. 

The natural frame to observe large-$p_\bot$ phenomena and to calculate the corresponding NLO cross section in deep-inelastic scattering (DIS) is the proton virtual-photon center-of-mass system (hadronic frame or CM$^*$). The large $p_\bot^*$ of the final photon provides the scale which can be compared with the photon virtuality; a large ratio $p_\bot^{*2}/Q^2$ defines the kinematical range in which the photon structure function formalism is useful. Such an approach, but without the introduction of the virtual photon structure function, can be found in \cite{d}. It contains detailed studies on the jet algorithms and the scale choice dependence of the cross sections.

As the kinematical conditions are often specified in the laboratory frame and as a large-$p_\bot$ in the laboratory does not necessarily implies a large $p_\bot^*$ in the CM$^*$, a lower limit $p_\bot ^* > E_{\bot cut}^*$ can also be imposed by the experiments. This condition will preserve the validity of a perturbative calculation and the possibility to define a virtual photon structure function. The production of jets and of forward $\pi^0$ has been measured with this convention by H1 \cite{7,8} and ZEUS \cite{9}. On the other hand, several experiments have also used the laboratory frame (LAB frame) to present their results \cite{a,b,c} without imposing the requirement $p_\bot^* > E_{\bot cut}^*$. As we shall see, the approach involving the definition of the resolved cross section is not always under control, and we have to content ourselves with the calculations of unsubtracted direct contribution. Thus we loose the possibility to supplement them with HO corrections. 

In this paper we consider DIS reactions in which an isolated photon and a jet are observed in the final state, extending the approach used in the inclusive case \cite{g} with only a photon observed in the final state. We discuss both cases, when the transverse momenta are defined in the CM$^*$ or in the LAB frames. This study is triggered by recent detailed ZEUS results \cite{c}. Unfortunately no $p_\bot^*$ constraint has been introduced by this collaboration, thus forbidding, in some kinematical domains, direct comparisons with complete NLO predictions.

The comparison with inclusive isolated cross section done in our previous paper was favored by the H1 kinematics \cite{a} having a large
domain where the condition $p_\bot ^{*2} > E_{\bot cut}^{*2}$ with $p_\bot ^{*2} /Q^2 \ \gsim \ 1$ was verified. The situation is less
favorable with the ZEUS kinematics having a larger range in $Q^2$. The observation of a photon and a jet in the laboratory does not necessarily
imply a large invariant mass squared $s_{\gamma j}$ of this system (when the photon and the jet are almost parallel). Therefore the addition of a
jet is not sufficient to prevent configuration with $p_\bot ^* < E_{\bot cut}^*$. This fact leads us to introduce a cut-off $E_{\bot cut}^*$ in our calculations and to study the stability of our results when $E_{\bot cut}^*$ goes to zero. If, in some kinematical domains, the results are not sensitive to $E_{\bot cut}^*$ we will compare them with ZEUS data.

The plan of this paper is the following. The next section is devoted to the calculation of the gamma-jet cross section in the CM$^*$. By this we mean that the isolation algorithm for the photon and jet algorithm are defined in this frame. The jet algorithm also contains conditions such as ``jet of highest $p_\bot$'' in the CM$^*$. As we fix the kinematical variable $p_\gamma$ and $p_{jet}$ in the laboratory we require the condition $p_{\bot \gamma}^* > E_{\bot cut}^*$. We discuss in detail the introduction of the resolved component. The four contributions (d-d, d-f, r-d and r-f) are calculated and their respective importance is discussed. Then, in section 3, we discuss calculations performed in the laboratory frame. By this we mean that the isolation algorithm and jet algorithm are now defined in the laboratory where the kinematical boundaries are also fixed.  We furthermore study the differences between the standard cone algorithm and the democratic one for the photon isolation and for the jet definition. We have to introduce a cut-off $E_{\bot cut}^*$ in order to avoid possible instability of the cross section. Finally section 4 is devoted to the comparison with the recent ZEUS data~\cite{c}.

\section{Proton Virtual-photon Center of Mass Frame}
\hspace*{\parindent}  
In this section, devoted to calculations in the CM$^*$, we study the four contributions to the $\gamma$-jet cross section discussed in the introduction. Such a frame looks like the laboratory frame of a hadronic collider; the colliding particles have collinear trajectories and the large transverse momentum $p_\bot^*$ of the observed final particle fixes the large scale appearing in the distribution functions and in the strong coupling constant. \par

However to be close to the experimental conditions we impose constraints on the observed kinematical variables, transverse momenta and rapidity, in the laboratory frame. The particle momenta are transformed to the CM$^*$ frame in which we define the isolation algorithm and the jet algorithm. As discussed in the introduction we impose a cut on the photon momentum $p_{\bot \gamma}^* > E_{\bot cut}^*$ in order to avoid instabilities of the cross section and to guarantee the validity of the perturbative regime. In this section we use the cone algorithm \cite{a)} for the isolation and the longitudinal $k_\bot$-algorithm \cite{b)} to define a jet. When two jets are present, we observe the jet of higher $p_\bot ^*$. For the photon the radius of the isolation cone is $R^* = 1.$ and we require a ratio ${E_\bot^{*had} \over E_\bot^{*\gamma} + E_\bot^{*had}}  \leq .1$ where $E_\bot^{*had}$ is the hadronic energy contained in the isolation cone. For the jet the radius $R_{jet}^*$ of the $k_\bot$-algorithm is set equal to $R_{jet}^* = 1$. Another algorithm, the so-called democratic algorithm will be discussed in section 3.\par

We use the CTEQ6M proton structure \cite{c)} and the virtual photon structure function presented in ref. \cite{f} and used in refs. \cite{g} and \cite{e)}.  The fragmentation function is that of the BFG collaboration (set II) \cite{h)}. Details on the NLO calculation can be found in ref. \cite{g,e),i),j)}. We work in the $\overline{\rm MS}$ scheme for factorization and renormalization with $\Lambda_{\overline{MS}}(4) = 236$~MeV and $N_f = 4$. The factorization and the renormalization scales are taken equal to $p_{\bot\gamma}^{*2} + Q^2$. The kinematical constraints are defined in the laboratory. We adopt those of ref.~\cite{c}. The colliding electron and proton have energies $E_e = 27.6$~GeV and $E_p = 920$~GeV corresponding to the center-of-mass energy $\sqrt{s} = 319$~GeV. The photon momentum has to lie within the ranges $4 < E_\bot^\gamma < 15$~GeV and $-.7 < \eta^\gamma < .9$. The jet momentum is required to have  2.5~GeV $< E_\bot^{jet} < $ 35~GeV and $-1.5 < \eta^{jet} < 1.8$. Constraints on the final electron are~: $E{'}_e > 10$~GeV, $\theta_e > 140^\circ$ (the $z$ axis is pointing toward the proton direction). Finally the photon virtuality is $10 < Q^2 < 350$~GeV$^2$. The numerical calculations are carried out using the adaptative Monte Carlo code BASIS~\cite{basis}.

\subsection{Direct-direct contribution}
\label{sec:dir-dir-CM}
\hspace*{\parindent} 
In the d-d contribution the real and virtual photons are directly coupled to the hard subprocess and the Born term, $\gamma^*\,q \rightarrow \gamma\,q$, is particularly simple: it consists of the Compton scattering diagrams.  The cross section has a ${1 / \widehat{t}}$ behavior\footnote{Mandelstam variables with a hat are those of the subprocess.} when the photon is emitted close to the forward direction (collinear to the initial quark) and a ${1 / \widehat{s}}$ behavior when it is emitted collinear to the final quark. In the CM$^*$ the divergent configurations, in $ \widehat{s}$ or $\widehat{t}$, are forbidden by the requirement $p_{\bot\gamma}^* > E_{\bot cut}^*$. \par

Real HO corrections to the Born term are given by amplitudes involving an extra gluon coupled to the quark line. When this gluon is collinear to the initial quark the cross section is divergent. This divergence is subtracted and absorbed in the quark distribution. When this gluon is emitted at large $p_\bot^*$, the initial virtual photon can produce a collinear quark-antiquark pair. This is the origin of the virtual photon structure function. After integration on the final quark phase space one gets the expression $\sigma_{q\bar{q} \to \gamma g}^B (Q^2, p_{\bot\gamma}^*, y_\gamma^*) \otimes P_{\bar{q}\gamma} \ln \left (  {p_{\bot\gamma}^{*2} + Q^2 \over Q^2}\right )$ where the symbol $\otimes$ indicates a convolution between the Born cross section and the antiquark distribution in the virtual photon\footnote{$z$-dependent factors in the logarithm argument are not indicated. They are discussed in ref. \cite{f}.}. $P_{\bar{q}\gamma}$ is the DGLAP branching function $P_{\bar{q}\gamma}(z) = {3 \alpha \over 2\pi}\ e^2 [(1-z)^2 + z^2]$. In the large-$p_\bot^{*}$ regime, ${p_{\bot\gamma}^{*2} \over Q^2} \gg 1$, we can neglect $Q^2$ in the Born cross section. In order to resum the structure function we subtract $\sigma^B(0, p_{\bot\gamma}^* , y_\gamma^*) \otimes P_{\bar{q}\gamma}  \ln \left({M_\gamma^2 \over Q^2}\right)$ from the d-d HO term thus defining the HO$_s$ corrections. The subtracted term builds, through the inhomogeneous DGLAP evolution equation, the resummed structure function $G_{\gamma^{*}}^{q} (z, Q^2, M_\gamma^2)$ \cite{f}, the basis of the resolved contributions. We choose $M_\gamma^2 = p_{\bot\gamma}^{*2} + Q^2$ so that the structure function vanishes when $p_{\bot\gamma}^{*2} \ll Q^2$, and is proportional to $\ln {p_{\bot\gamma}^{*2} \over Q^2}$ for ${p_{\bot\gamma}^{*2} \over Q^2} \gg 1$.\par

\begin{figure}[t]
\centering
\null\vspace{-0.2cm}
\includegraphics[scale=.75]{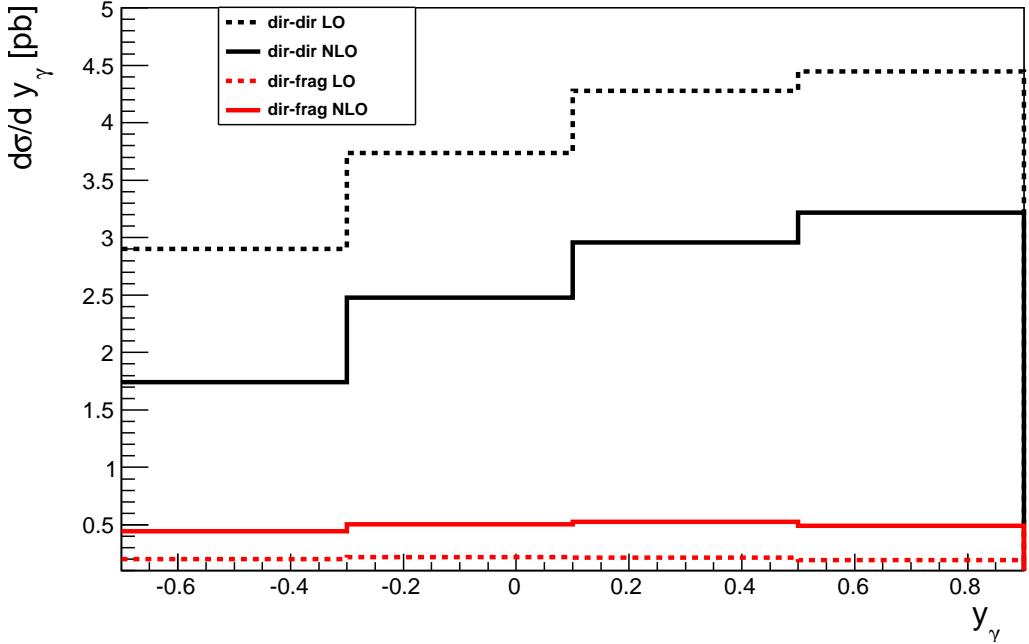}
\caption{\it The photon rapidity distribution for the direct-direct (upper two curves, black) and direct-fragmented (lower two curves, colored) terms at the leading order and the next-to-leading order. The NLO cross sections are understood with the $P_{\bar{q}\gamma}  \ln \left( {M_\gamma^2 \over Q^2}\right)$ term subtracted and the cut $E_{\bot cut}^* = 2.5~\mbox{GeV}$ is implemented (see text).}
\label{fig:ygam-dir-CM}
\end{figure}

Some numerical results are displayed in Fig.~\ref{fig:ygam-dir-CM} for the cross section $d\sigma^{d-d}/dy_\gamma$ with the cut $E_{\bot cut}^* = 2.5$~GeV. The kinematics has been 
given at the beginning of section 2; $y_\gamma$ is the rapidity in the laboratory. As expected the cross section is largest in the forward direction due to the dominance of the $\widehat t$ term in the scattering amplitude. \par

We notice that the HO$_{s}$ corrections are negative, as already
observed in the inclusive case \cite{g}. On the contrary, the unsubtracted HO corrections are close to zero, the difference HO - HO$_{s}$ being the contribution of the lowest order resolved term that will be discussed in the next subsection. In fact, the NLO unsubtracted result cannot be distinguished from the d-d LO curve in Fig.~\ref{fig:ygam-dir-CM}.  The results in that figure strongly depend on the value of the cut-off $E_{\bot cut}^*$. This point is illustrated in table 1 where the total $\overline{\rm MS}$ cross section is displayed. We consider two domains in $Q^2$.
We see that for low value of $Q^2$ the NLO cross section is stable with respect to the $E_{\bot cut}^*$ value while it is not the case when considering the whole $Q^2$ range. In the former case, the Lorentz boost between the laboratory and the CM$^*$ frames, $q_\perp = \sqrt{Q^2(1-y)}$ with $y$ the usual DIS variable (scaled virtual-photon energy in the laboratory), is small and  $p_{\perp\gamma}^*$ remains relatively large, and close to its $p_{\perp\gamma}$ value in the laboratory, so that the cross section is insensitive to the cut-off. On the contrary for large values of $Q^2$,   $p_{\perp\gamma}^*$ may reach small values where perturbation theory is not reliable (very large HO corrections) and where the theoretical predictions become sensitive to the cut-off. The errors associated to the NLO cross sections are the values resulting from the Monte-Carlo integration procedure (for the Born cross sections the errors are at the per-mil level). 

\begin{table}
\begin{center}
\begin{tabular}{|c|c|c|c|c|}
\hline
 &\multicolumn{2}{|c|}{} &\multicolumn{2}{|c|}{}  \\
&\multicolumn{2}{|c|}{10 $< Q^2< $ 350 GeV$^2$} &\multicolumn{2}{|l|}{\qquad 10 $< Q^2< $ 50 GeV$^2$}  \\
&\multicolumn{2}{|c|}{} &\multicolumn{2}{|l|}{}  \\
\hline
&&&&\\
$E_{\bot cut}^*$ &{Born}  &{NLO}&{Born} &{NLO} \\
&&&&\\
\hline
&&&&\\
1.5 GeV  &7.50 &$2.16 \pm  .05$ &3.51 & $2.99 \pm .03$\\
&&&&\\
\hline
&&&&\\
2.5 GeV  &$6.14$ &$4.21 \pm .05$&3.39 &$2.91 \pm .03$ \\
&&&&\\
\hline
\end{tabular}
\end{center}
\caption{\it Variation of the d-d $\gamma$-jet cross section in pbarns with $E_{\bot cut}^*$ in two $Q^2$ ranges. The isolation and jet criteria are implemented in the CM$^*$.}
\label{table1}
\end{table}

\subsection{Resolved-direct contribution}
\hspace*{\parindent}
The r-d contribution that we consider in this subsection is made up of two terms, the Born cross section and the HO
corrections to the latter. Here the Born cross section is given by the convolution of the parton distributions in the virtual
photon with Born subprocesses. In the preceding subsection we found the lowest order expression of this distribution. Here we
use the resummed expression calculated at NLO is the $\overline{\rm MS}$ factorization scheme~\cite{f}. We can compare the resummed
Born contribution of this subsection $\sigma^{Born}$(r-d) with the lowest order one by considering the unsubtracted  HO
contribution and the subtracted $d$-$d$ contribution HO$_s$ found in the preceeding subsection (integrated over $y_\gamma$). 

$$\begin{array}{ll}
\sigma^{HO} - \sigma^{{HO}_s} &= 6.13 - 4.21 = 1.92\ \mbox{pb}\\
\sigma^{Born} \hbox{(r-d)} &=  2.12\ \mbox{pb}\\
\end{array}
$$

\noi The slight difference can be attributed to the difference between the lowest order $\gamma^*$ structure function and the resummed one (which also contains gluons). HO corrections for the r-d contribution can be borrowed from prompt photon production in hadronic collision, by using the code JETPHOX \cite{k)} in which a hadronic parton distribution is replaced by the virtual photon distribution function. We find a large effect $(E_{\bot cut}^* = 2.5$~GeV)~:
$$\begin{array}{l}
\hbox{$\sigma^{Born}$(r-d) = 2.12}\ \mbox{pb}\\
\\
\hbox{$\sigma^{NLO}$(r-d) = 3.30}\ \mbox{pb}\\
\end{array}$$
\noi which illustrates the interest to consider the HO corrections in the r-d contribution. Here also the $E_{\bot cut}^*$ dependence is large. For $E_{\bot cut}^* = 1.5$ GeV we obtain $\sigma^{Born}$ = 5.04 pb and $\sigma^{NLO}$ = 7.24\ {pb}.\par

\begin{figure}[h]
\centering
\includegraphics[scale=.75]{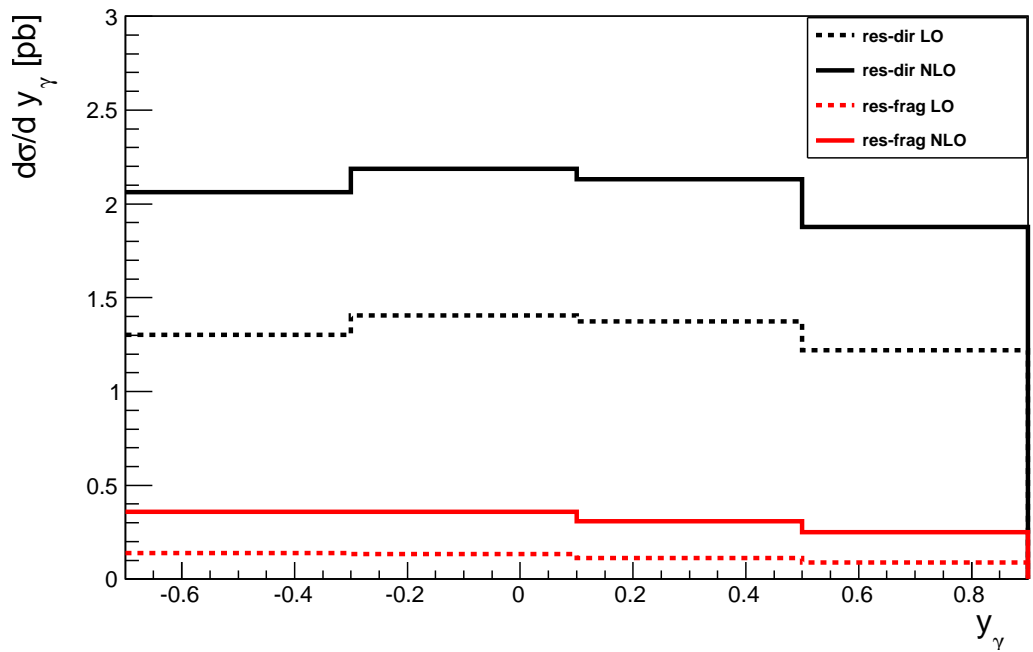}
\caption{\it The rapidity distribution for the resolved-direct (upper two curves, black) and resolved-fragmented (lower two curves, colored) terms at the leading order and the next-to-leading order}
\label{fig:ygam-res-CM*}
\end{figure}

The Born and NLO cross sections as function of $y_\gamma$ are displayed in Fig. \ref{fig:ygam-res-CM*}. Unlike in the d-d case the cross section is rather flat in rapidity and even slightly decreasing at large rapidity: this is understood because the photon can now also be emitted by a quark in the $\gamma^*$, $i.e.$ in the backward direction. Concerning the HO corrections we see that they are not negligible.

\subsection{Direct-fragmentation contribution}
\hspace*{\parindent}
The direct-fragmentation contribution, obtained with the BGF fragmentation functions, is considerably reduced by the isolation requirement: no parton in the photon cone carrying more than 10$\%$ of the total photon $+$ parton energy. It is roughly 15\% of the inclusive d-f cross section. We obtain, always with $E_{\bot cut}^* = 2.5$~GeV, the following results
$$\begin{array}{l}
\hbox{$\sigma^{Born}$(d-f) = .33 pb}\\
\\
\hbox{$\sigma^{NLO}$(d-f) = .78$\ \pm .01\ $} \mbox{pb}\\
\end{array}$$

\noindent which are one order of magnitude smaller than the direct-direct contribution. The $y_\gamma$ rapidity spectrum is shown is Fig. \ref{fig:ygam-dir-CM}. It is relatively flat. This is related to the fact that the photon can now be radiated by a parton in the backward direction ($\widehat u$ channel pole). This new channel which populates a different region of phase-space, compared to the d-d term, receives large HO corrections associated to diagrams involving the triple gluon coupling. Not unexpectedly, we observed the same behavior as for the inclusive cross section \cite{g}. 

\subsection{Resolved-fragmentation contribution}
\hspace*{\parindent}
The r-f contribution is, after isolation, the smallest of the four contributions. We have:
$$\begin{array}{l}
\hbox{$\sigma^{Born}$(r-f) = .19 pb}\\
\\
\hbox{$\sigma^{NLO}$(r-f) = .50$\ \pm .04\ $} \mbox{pb}\\
\end{array}$$
The photon rapidity distribtion is shown in Fig. \ref{fig:ygam-res-CM*}. The large HO corrections are due to many new processes. This process is in fact similar to a pure hadronic one such as $p + p \rightarrow \pi^0 + jet + X$ which is known to receive large higher order corrections \cite{k1)}.

\section{Laboratory  Frame}
\hspace*{\parindent}
The definition of the photon-jet cross section in the laboratory frame
introduces several new features not present previously. All these features come from the fact that a large-$p_\bot$ parton in the laboratory does not necessarily correspond to a large-$p_\bot$ parton in the CM$^*$. Therefore the simple picture of two initial collinear partons scattering into two large-$p_\bot$ final partons and of the associated perturbative HO corrections may break down. The spectator jets, for instance those of the resolved photon, which are low-$p_\bot$ jets in the CM$^*$ can
appear as large-$p_\bot$ jets in the laboratory , and must be included in the
definition of the photon-jet cross section. The jet hierarchy also can be
changed; the highest-$p_\bot$ jet in the CM$^*$ does not necessarily correspond to the highest one in the laboratory. \par

A more adapted approach for a laboratory frame calculation is that of \cite{e}
involving no $E_{\bot cut}^*$. But this approach leads to a new singularity in the
perturbative calculations, that of the final collinear contribution of the
subprocess $\gamma^* + q \to q + \gamma$. Therefore at lowest $O(\alpha^2)$  order this  calculation requires the introduction of a fragmentation function to absorb the
final state singularity. At present there is no $O(\alpha^2 \alpha_s)$ calculation available analogous to that performed in the $\gamma^* \to q+\overline{q} + \gamma$ channel  \cite{l)}. Therefore we continue our CM$^*$ approach adapted to the laboratory frame which implies a cut $E_{\bot cut}^*$. \par
 
In this section we study the photon-jet cross section measured in the laboratory. This means that isolation and jet algorithms are defined in terms of parton momenta in the laboratory frame. The jet is the jet of highest-$p_{\bot}$ in the laboratory. Moreover if the highest-$p_{\bot}$ jet is outside the acceptance in rapidity we take into account the smaller one. For each of the four topologies there are specific features that we discuss in the following subsections.

\subsection{Direct-direct contribution}
\hspace*{\parindent}
We start this section by discussing two algorithms used to define isolated
photons. They are the cone algorithm, already used in the CM$^*$ calculation, and the democratic $k_\perp$-algorithm \cite{glover} used, for instance, in the ZEUS experiment \cite{c}. For the definition of jets we use the longitudinal $k_\perp$ algorithm~\cite{b)}. All momenta are measured in the laboratory.\par

In our case we have at most two partons besides the photon in the final state~: in the direct-direct case the photon is (in the theoretical sense) isolated while in the direct-fragmented case it is accompanied by a collinear parton. The cone algorithm is the simplest one to implement. One calculates the distance between the photon and each parton ($i = 1,2$)~: 
\beq
\label{1e}
R_{\gamma i} = \sqrt{\delta \eta_{\gamma i}^2 + \delta \phi_{\gamma i}^2} \ ,
\eeq
where the quantities under the $\sqrt{}$ are respectively the difference in (pseudo-)rapidities and in azimuthal angles of the photon and parton $i$. Then if $R_{\gamma i} < 1$, the parton is in the photon cone and we have to test if the total transverse hadronic energy in the cone $E_{\bot i}$ (including also in the d-f case the energy of the fragment collinear to the photon)
\beq
\label{2e}
E_{\bot i} < \varepsilon \left ( E_{\bot i} + E_{\bot \gamma} \right ) 
\eeq
with $\varepsilon = .1$ as in the case of ZEUS experiment. If both partons are outside the photon cone, then one tests if they form one or two jets depending on whether their distance $R_{12} =  \sqrt{\delta \eta_{12}^2 + \delta \phi_{12}^2}$ is smaller or larger than $1.$ If $R_{12} < 1$ they form one jet with coordinates:
\bea
\label{3e}
E_{\bot jet} &=& E_{\bot i} + E_{\bot j} \nn \\
\eta_{jet} &=& \left ( E_{\bot i} \ \eta_i + E_{\bot j} \ \eta_j \right )/E_{\bot jet}\nn \\
\phi_{jet} &=& \left ( E_{\bot i} \ \phi_i + E_{\bot j} \ \phi_j \right )/E_{\bot jet} \ .
\eea
If this jet falls within the detector acceptance we keep the event. In the case of two jets,  we test for the small $E_\bot$  jet if the largest one does not fall in the acceptance. It may also happen that both partons satisfy the criterium $R_{\gamma i}  < 1$ in which case we, of course, drop the event since there is no visible jet in that case.\par

In the democratic-$k_\bot$ algorithm case we treat the photon as an ordinary parton $i = \gamma , 1, 2$ and we order the distances defined in the following way~:
\beq
\label{4e}
d_i = E_{\bot i} \ , \quad d_{ij} = \ {\rm min} \left ( E_{\bot i}, E_{\bot j}\right ) \sqrt{\delta \eta_{ij}^2 + \delta \phi_{ij}^2} /R \ ,
\eeq

\noi with $R = 1$ as in Zeus case. If the smallest of these variables is a $d_i$ we remove it and call it a jet or an isolated photon, provided that, in the fragmentation case the parton collinear to the photon satisfies the condition (\ref{2e}). We then continue the procedure with the remaining  distances. If the smallest variable is a $d_{ij}$ with $i, j = 1, 2$ we combine the partons to form a jet with the coordinates as in eq. (\ref{3e}) and we perform, with this jet, the isolation tests as above. If the smallest distance is a $d_{\gamma i}$, $i = 1,2$, we test for the isolation condition (\ref{2e})~: if it is satisfied we calculate the coordinates of this ``isolated photon jet'' as in (\ref{3e}) and test whether or not the remaining jet falls within its cone.\par

The two algorithms should lead to very similar results~\cite{d}. In fact, in case there are only two large $E_\bot$ objects they are identical. However when there are three or more objects there appears some differences. For example, consider the hierarchy $d_{12} < d_{\gamma 1} < d_\gamma < d_{\gamma 2} , d_1, d_2$. In the cone algorithm, parton 1 is in the photon cone and the photon appears non-isolated because $E_{\bot\gamma} < E_{\bot 1}$, thus the energy fraction criterium is not satisfied~: the event would be rejected. In the $k_\bot$ algorithm the two partons would form a jet and the photon would appear isolated. Such occurences are rare however. In the exemple above, the conditions $d_{12} < d_{\gamma 1} <  d_\gamma $ mean that the photon and both partons are "close" to one another which is possible only if there is a large transverse boost from the CM$^*$ frame to the LAB one.\par

\begin{table}
\begin{center}
\begin{tabular}{|l|c|c|}
\hline
&&\\
&Cone algor. &Democratic algor. \\
&& \\
\hline
&& \\
{Born} &6.44 &6.44 \\
&& \\
\hline
&& \\
{NLO} &$4.22 \pm .03$ &$4.35 \pm .03$ \\
&& \\
\hline
\end{tabular}
\end{center}
\caption{\it{The d-d $\gamma$-jet cross section calculated with the cone and the democratic algorithms, in pbarns. The isolation and jet criteria are implemented in the laboratory.}}
\label{table2}
\end{table}
We obtain quasi identical results for the cross sections calculated with the two algorithms. They are displayed in table 2. We used the parameters and proton distribution functions defined in section 2. 
Unlike in the analysis in the CM$^*$ of sub-section \ref{sec:dir-dir-CM} the d-d NLO cross section is insensitive to the $E_{\bot cut}^*$ cut-off :
the cross section is reduced by 1\% to 2.5\% at negative $y_\gamma$ and unchanged at positive $y_\gamma$ when varying the cut-off from 0 to 2.5 GeV. The fact that the cross sections are not divergent is due to the photon isolation and the requirement of a jet in addition to the isolated photon.
Therefore in the direct-direct case we are able to take the limit $E_{\bot cut}^* \to 0$ since the experimental requirements forbid the final collinear $q - \gamma$ configuration. This independence of the cross section on the cut-off means that the ZEUS cuts are sufficient to constrain the d-d term to remain in the perturbative regime where the HO$_{s}$ corrections to the Born term are under control.\par
\begin{figure}[b!]
\centering
\includegraphics[scale=.75]{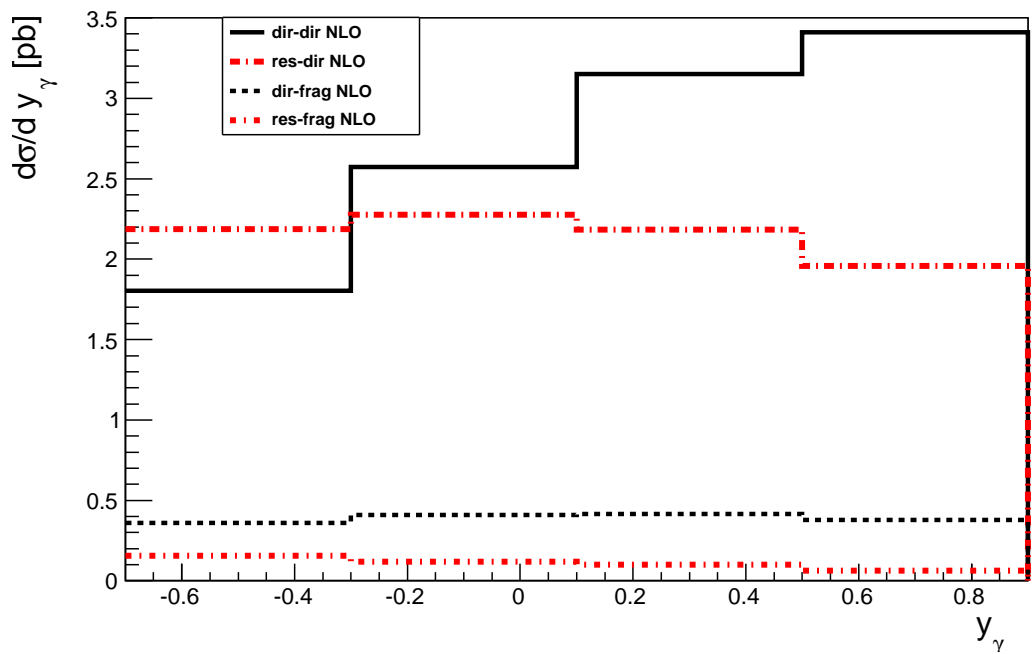}
\caption{\it The photon rapidity distribution for the direct and resolved  terms at the next-to-leading order, with photon isolation and jets defined in the laboratory; the lower two curves are the fragmentation components. For all components the constraint $E^*_{\perp cut} = 2.5$ GeV is implemented.}
\label{fig:ygam-res-LAB}
\end{figure}

In Fig. \ref{fig:ygam-res-LAB} we display the d-d contribution at NLO (solid line) to the cross section $d\sigma^{LAB}/dy_\gamma$ calculated with $E_{\bot cut}^* = 2.5$ GeV and the democratic $k_\bot$- algorithm. \\

Let us note that the laboratory isolation criteria, without imposing an $E_{\bot cut}^*$ condition, lead to a cross section very similar that of table 1 where the constraint $E_{\bot cut}^*> 2.5$ GeV was imposed and jet and isolation defined in the center of mass frame (a cross section of 4.21 $\pm$ 0.05 pb was found in that case). \par

\subsection{Resolved-direct contribution}
\hspace*{\parindent}
The structure of the resolved-direct contribution defined in the laboratory frame is more complicated than the direct-direct one. The complication arises from the fragments of the $\gamma^*$ structure function. These fragments may go into the final photon isolation cone or may be seen as a jet. Moreover, in the higher order contributions we also have configurations with the low $p^*_\perp$ photon in the CM$^*$ collinear to a parton of the $\gamma^*$ structure function, which appears at large transverse momentum in the laboratory. This configuration is divergent (in the collinear limit) and is forbidden in the CM$^*$ frame by the fact that only large $p^*_\perp$ photons are observed. In the laboratory this configuration is forbidden by the condition $E_{\bot cut}^* > 0$. Therefore we expect a strong dependence on $E_{\bot cut}^*$ of the resolved-direct contribution. This dependence is shown in table 3.
\begin{table}[b!]
\begin{center}
\begin{tabular}{|c|c|c|c|c|}
\hline
 &\multicolumn{2}{|c|}{} &\multicolumn{2}{|c|}{}  \\
$E_{\bot cut}^*$  [GeV] &\multicolumn{2}{|c|}{$10 < Q^2 < 350$ GeV$^2$} &\multicolumn{2}{|l|}{$10 < Q^2 < 50$ GeV$^2$}  \\
 &\multicolumn{2}{|c|}{} &\multicolumn{2}{|c|}{}  \\
\cline{2-5}
&&&&\\
 &Born &NLO &Born &NLO \\
&&&&\\
\hline
&&&&\\
.5 &2.07 &5.02 &.96 &2.17  \\
&&&&\\
\hline
&&&&\\
1.5 &2.06 &4.41 &.96 &2.06\\
&&&&\\
\hline
&&&&\\
2.5 &1.70 &3.40 &.92 &1.81  \\
&&&&\\
\hline
\end{tabular}
\end{center}
\caption{\it{Variation with $E_{\bot cut}^*$ of the r-d $\gamma$-jet cross section in pbarns, in various phase space domains. The isolation and jet criteria are implemented in the laboratory.}}
\label{table3}
\end{table}
\noindent 
However we note that this dependence is weaker when $Q^2 < 50$~GeV$^2$. With this kinematical requirement we obtain a more stable cross section. \par

The r-d contribution at NLO (long-dash-dotted line) to the cross section $d\sigma^{LAB}/dy_\gamma$ is displayed in Fig.~\ref{fig:ygam-res-LAB} for $E_{\bot cut}^* = 2.5$~GeV: it has the same features as in Fig. \ref{fig:ygam-res-CM*}, namely very slightly decreasing with $y_{\gamma}$ and one finds large HO corrections throughout the whole range (about a factor 2).

\subsection{Direct-fragmentation contribution}
\hspace*{\parindent}
The $y_\gamma$ distribution of the d-f component, with $E_{\bot cut}^* = 2.5$ GeV, is shown in Fig.~\ref{fig:ygam-res-LAB}. It is essentially flat and rather small, the same features as in Fig.~\ref{fig:ygam-res-CM*}. The HO contribution is roughly independent of $y_\gamma$ and doubles the Born term. The distribution is relatively stable under the $E_{\perp cut}$ parameter: if taken to zero the d-f component is 13\% larger in the backward bin decreasing to no variation in the most forward bin.

\subsection{Resolved-fragmentation contribution}
\hspace*{\parindent}
We have in the r-f HO contribution a piece similar to that in the r-d HO one. A large-$p_\bot$ parton (in the LAB) can be emitted collinearly with an initial parton in the resolved virtual photon and then fragments into a photon. Again this corresponds to a collinear configuration in the CM$^*$ frame that we cut by the requirement $E_{\bot cut}^* > 0$. Therefore this contribution has a strong dependence on this cut-off parameter.
However, as seen in Fig. \ref{fig:ygam-res-LAB} it is the smallest of all four components and will give a negligible contribution to the full isolated $\gamma$-jet cross section.

\section{Comparison to ZEUS data}
\begin{figure}[!t]
\centering
\null\vspace{-2.cm}
\includegraphics[scale=.75]{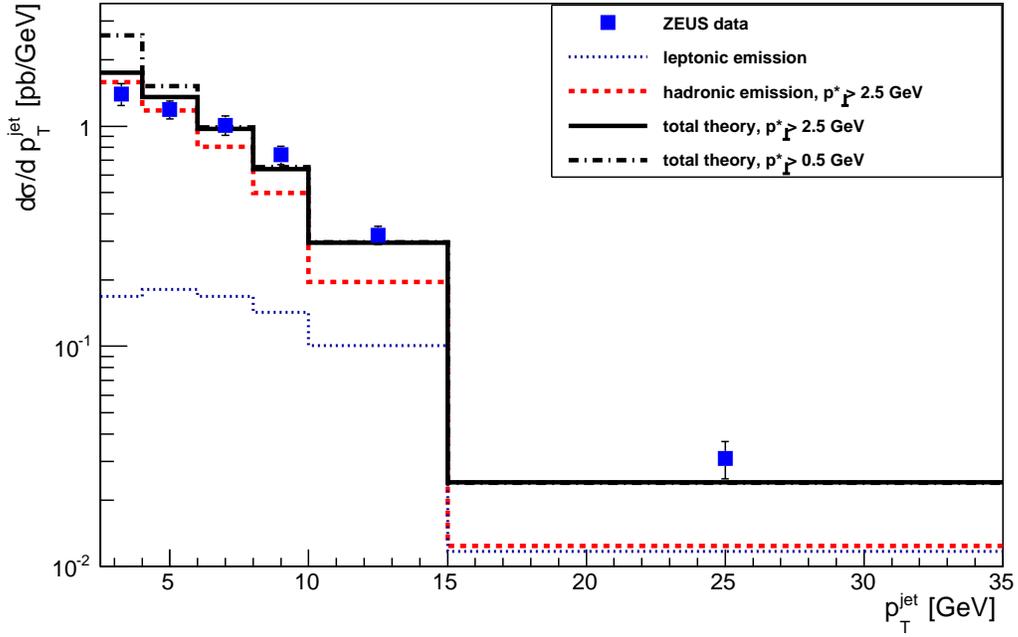}
\caption{\it The $p_\perp^{jet}$ distribution: comparison of the full NLO calculation to ZEUS data  \cite{c}. The leptonic contribution is as quoted in \cite{c}.}
\label{fig:PTjet-ZEUS-LAB}
\end{figure}

\begin{figure}[!h]
\centering
\includegraphics[scale=.75]{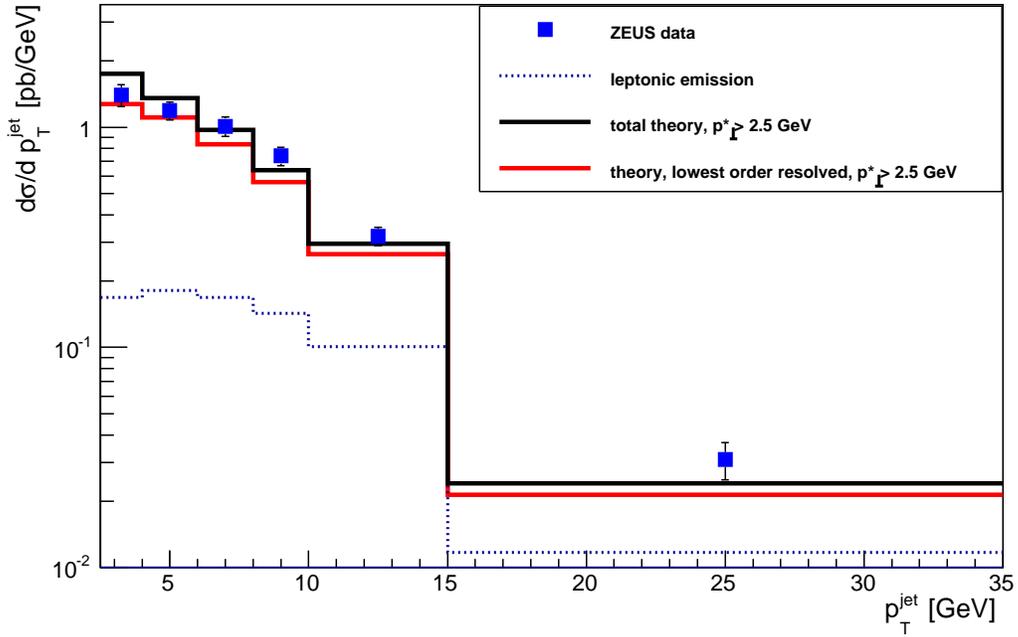}
\caption{\it The $p_\perp^{jet}$ distribution: comparison of the partial NLO calculation, with the unresummed resolved component, to ZEUS data. The data and the leptonic contribution are as quoted in \cite{c}.}
\label{fig:PTjet-ZEUS-NORES-LAB}
\end{figure}

After this lengthy discussion on each of the components which build the $\gamma$-jet cross section we are now ready to compare the NLO calculation to the most recent ZEUS data. It is worth stressing that only the sum of all 
\begin{figure}[!t]
\centering
\null\vspace{-2.cm}
\includegraphics[scale=.75]{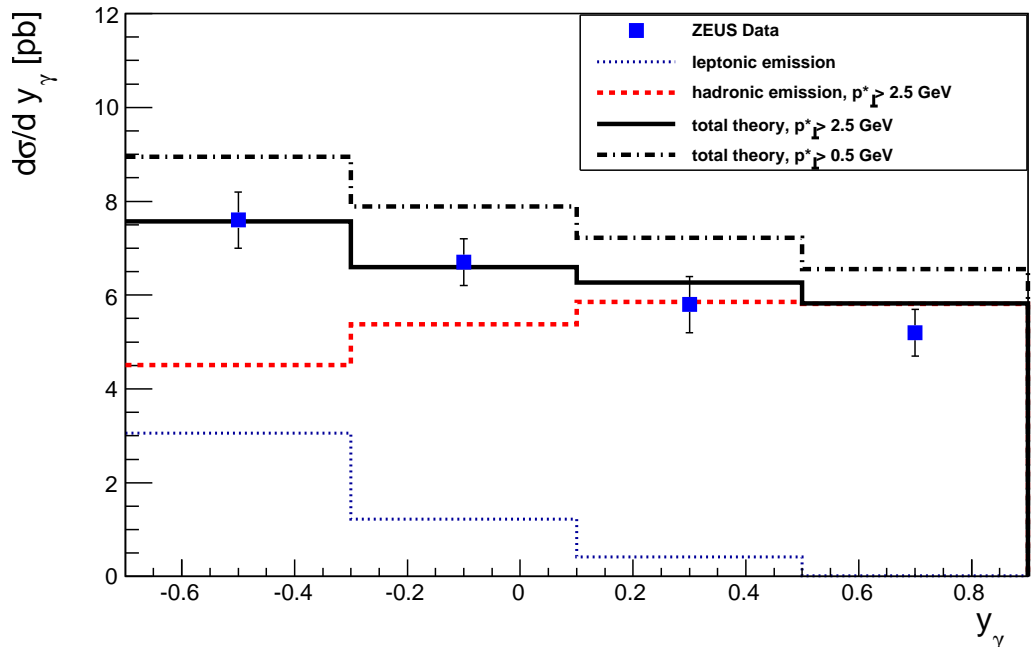}
\caption{\it The photon rapidity distribution: comparison of the full NLO calculation to ZEUS data  \cite{c}. The leptonic contribution is as quoted in \cite{c}.}
\label{fig:ygam-ZEUS-LAB}
\end{figure}
\begin{figure}[!h]
\centering
\includegraphics[scale=.75]{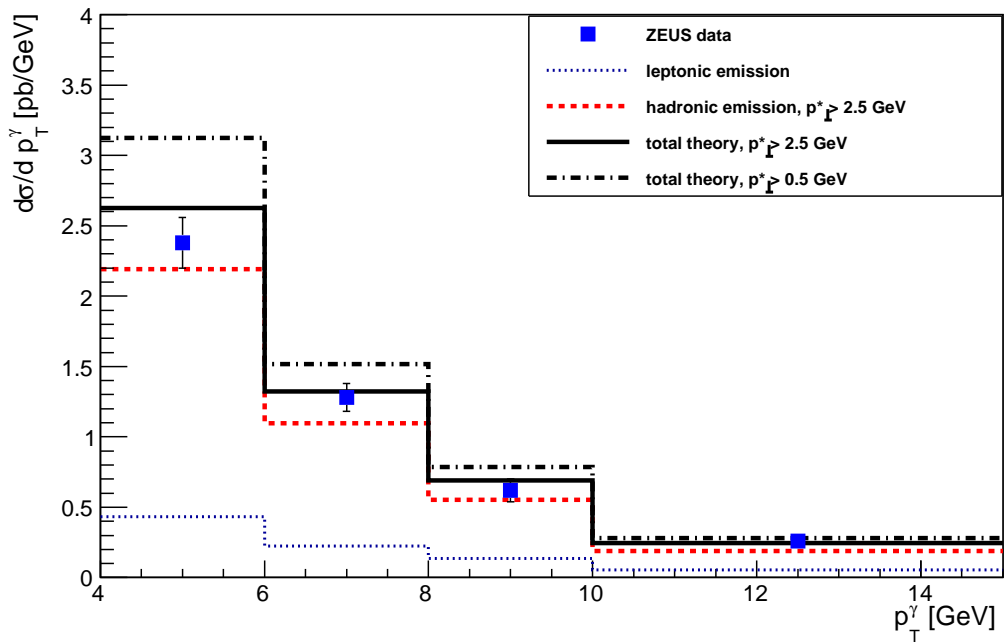}
\caption{\it The photon transverse momentum distribution: comparison of the full NLO calculation to ZEUS data  \cite{c}. The leptonic contribution is as quoted in \cite{c}.}
\label{fig:PT-ZEUS-LAB}
\end{figure}
four components have a physical meaning and can be compared to data as the relative  weight of each component depends on the unphysical scales. It has been stressed above that measuring the transverse components in the laboratory frame rather than the CM$^*$ frame does not garantee the perturbative stability of the calculation: indeed it appears from the above discussion that, if the direct components are rather stable under the $E^*_{\perp cut}$ parameter, this is not the case for the resolved ones because the ZEUS phase space does not garantee that the CM$^*$ momenta in the hard subprocesses remain large and also does not eliminate all the collinear singularities associated to the photon radiation from a quark at large laboratory transverse momentum. For these reasons we will bracket the theoretical predictions 
\begin{figure}[!t]
\centering
\null\vspace{-2.cm}
\includegraphics[scale=.75]{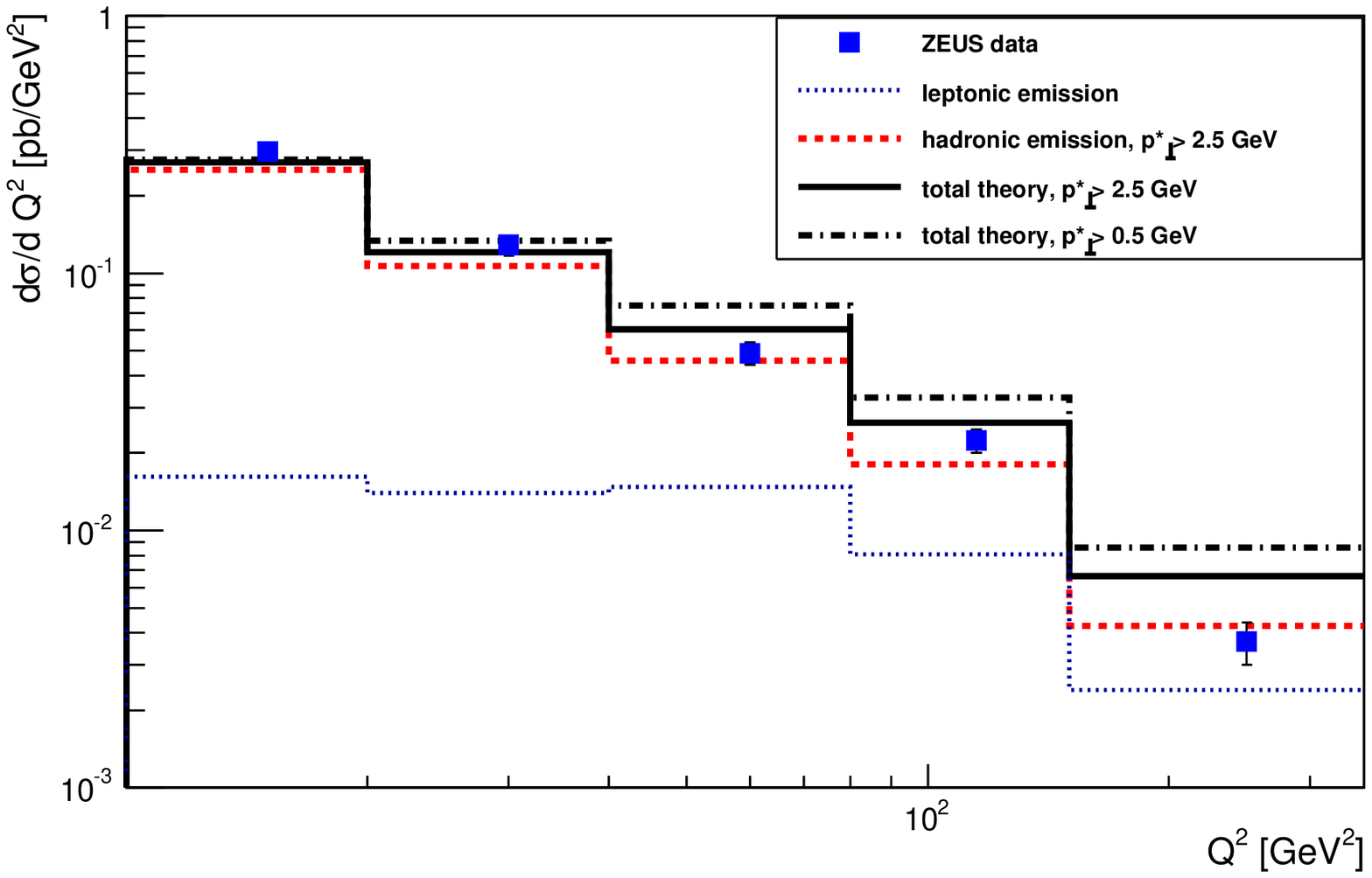}
\caption{\it The $Q^2$ distribution: comparison of the full NLO calculation to ZEUS data  \cite{c}. The leptonic contribution is as quoted in \cite{c}.}
\label{fig:Q2-ZEUS-LAB}
\end{figure}
by two cross sections: one obtained with $E^*_{\perp cut} = 2.5$ GeV, which garantees the perturbative stability but excludes some $\gamma$-jet configurations included in the data, the other with $E^*_{\perp cut} = 0.5$ GeV which accounts for a larger phase space at the cost of letting the theoretical predictions err in the non-perturbative regime.
The results of our approach are compared with data in Figs. \ref{fig:ygam-ZEUS-LAB}-\ref{fig:PTjet-ZEUS-NORES-LAB}. In all plots we include the contribution from the leptonic radiation of the photon ($e\rightarrow e + \gamma$) based on the DJANGOH~\cite{djangoh} generator, as estimated in \cite{c}. We add it to the hadronic results obtained in this work, with a numerical accuracy of better than 2\%, to obtain cross sections comparable to the data. The interference contribution between leptonic and hadronic emissions is negligible~\cite{c}. Furthermore no hadronization corrections are applied to our estimates. \\

In Fig. \ref{fig:ygam-ZEUS-LAB}  we display the photon rapidity spectrum and we note the very good agreement with the data when the cutoff $E^*_{\perp cut} = 2.5$ GeV is selected. Using the lower cut-off uniformely increases the predictions by about 12\%. In contrast, the predictions of \cite{d}, as quoted in ref. \cite{c}, where the resolved component is treated at lowest order, systematically underestimate the photon rapidity spectrum by about 20\%. \\


The same pattern is found in Fig. \ref{fig:PT-ZEUS-LAB} displaying the photon transverse momentum spectrum: very good agreement with $E^*_{\perp cut} = 2.5$ GeV, overestimate with $E^*_{\perp cut} = 0.5$ GeV. The interesting feature however is that both cut-offs lead to similar results in the high $p_\perp^\gamma$ tail where we expect the NLO resummed treatment of the resolved component to be most appropriate ($p_\perp^{*\gamma} \ge \sqrt{Q^2}$). There, our predictions can be taken as solid predictions from NLO perturbative QCD as all relevant scales are large in the CM$^*$ frame. Comparing to the estimates based on \cite{d} and quoted in \cite{c} we are above~: in fact they systematically underestimate the data while still being compatible with them at large $p_\perp^\gamma$.  \\
 
Turning to the $Q^2$ distribution in Fig. \ref{fig:Q2-ZEUS-LAB}, our results confirm the discussion above: excellent agreement and stability under a change of the cut-off at small $Q^2$, overestimate of the data at the larger values. Finally we conclude this phenomenological discussion with the $p_\perp^{jet}$ spectrum, in Fig. \ref{fig:PTjet-ZEUS-LAB}, where 
we notice a small excess
 in the predictions for the smallest two bins, $p_\perp^{jet} < 6$ GeV, with a large instability under change of cut-off in the first bin. In relation with this observation one notes a large higher correction (up to a factor 2 compared to the Born term)
to the resolved component. 
The origin of this is the fact that the collinear fragments of the virtual photon, which would be rejected by a cut-off in the CM$^*$ frame  can be seen as jets in the laboratory. In contrast, for  $p_\perp^{jet} > 6$~GeV we observe good agreement with data as well as a very good stability under change of the cut-off which indicates that NLO perturbation theory is valid. To further explore the small $p_\perp^{jet}$ regime we calculate the cross section without resummation on the resolved component (see Fig. \ref{fig:PTjet-ZEUS-NORES-LAB})~: agreement with data is achieved for the first two bins while above 6 GeV the cross section is systematically underestimated. A last remark concerning the comparison between theory and experiment at small $p_\perp$ is of a more practical nature~: the theoretical jets constructed at the partonic level are compared with the experimental ones at the hadronic level and  hadronization corrections not taken into account at NLO QCD must be important.

\section{Conclusions}
We have discussed at length the full NLO QCD predictions for $\gamma$-jet final states in deep-inelastic scattering at HERA and compared with ZEUS data where transverse momenta are defined in the laboratory frame. The motivation for studying these final states is to avoid the photon-quark final state collinear singularity  present in the Born term $\gamma^*\,q\rightarrow \gamma\,q$ when the photon, produced longitudinally in the CM$^*$ frame, is detected at large transverse momentum in the laboratory frame due to the transverse boost of the virtual photon. We have seen that the full NLO QCD calculation, taking into account the HO corrections to the resummed resolved component, again hits this collinear singularity problem which occurs in the large $Q^2$, low 
$p_\perp^\gamma$ or $p_\perp^{jet}$ regions. So, in deep inelastic scattering, doing large $p_\perp$ phenomenology with momenta defined in the laboratory is not enough to cure this collinear problem since it re-enters through the back door when considering higher order diagrams as are needed for precision calculations. Of course, working with CM$^*$ coordinates, or with a cut-off in the CM$^*$ frame, one would not have encountered this difficulty. 
Nevertheless, in some well identified kinematical regions in the laboratory (low $Q^2$ or large $p_\perp$) comparaison between experiment and the full NLO theory is astonishingly good unlike the case when considering partial NLO calculation where the resolved component is treated at lowest order. Of course further data involving a cut in the CM$^*$ frame should permit a sounder comparison between theory and experiment.
To be complete a study of the scale dependence of the theoretical estimates should have been done. However in view of the problems discussed  above this is not justified at this stage. It is enough to remember that the choice of scales used in this work ($M^2 = Q^2+p_T^{*2}$) gave very good agreement with data for related processes, namely $\pi^0$ production in DIS~\cite{e)} and photoproduction ($Q^2=0$) of isolated photons~\cite{zeus-photpro} (see \cite{fgh} for the theoretical discussion).
In this respect, we are not as pessimistic as the authors of \cite{dis-review} who observe, based on comparison of older data with previous theoretical calculations,  that isolated photon production both in DIS and photoproduction is underestimated in the NLO QCD approach.

\section{Acknowledgement}
We would like to thank Peter Bussey for carefully reading the manuscript and for his useful remark on the $p_T^{jet}$ spectrum.

\end{document}